\documentclass[12pt]{JHEP3}
\usepackage{epsfig}
\usepackage{amssymb}
\usepackage{bm}
\usepackage{amsmath}
\newcommand{\p}{\partial}

\makeatletter
\@addtoreset{equation}{section}
\makeatother

\title{
A Matrix Model for Baryons and \\Nuclear Forces
}
\author{
Koji Hashimoto$^{*,a}$,
Norihiro Iizuka$^{\dagger,b}$ and
Piljin Yi$^{\S,c}$\\
${^*}$ {\it Nishina Center, RIKEN, Saitama 351-0198, Japan}\\
${}^\dagger$
{\it Theory Division, CERN, CH-1211 Geneva 23, Switzerland}\\
${}^\ddagger$ {\it School of Physics, Korea Institute for Advanced
Study,
Seoul 130-722, Korea}\\
$^a$ E-mail: \email{koji(at)riken.jp}\\
$^b$ E-mail: \email{norihiro.iizuka(at)cern.ch}\\
$^c$ E-mail: \email{piljin(at)kias.re.kr}\\
}

\abstract{
We propose a new matrix model describing multi-baryon
systems. We derive the action from open string theory
on the wrapped baryon vertex D-branes embedded in the D4-D8
model of large $N_c$ holographic QCD. The positions of $k$
baryons are unified into $k\times k$ matrices, with spin/isospin
of the baryons encoded in a set of $k$-vectors. Holographic
baryons are known to be very small in the large 't Hooft coupling
limit, and our model offers a better systematic approach to
dynamics of such baryons at short distances. We compute
energetics and spectra ($k=1$), and also short-distance
nuclear force ($k=2$). In particular, we obtain a new
size of the holographic baryon and find a precise form of
the repulsive core of nucleons. This matrix model complements
the instanton soliton picture of holographic baryons, whose
small size turned out to be well below the natural length
scale of the approximation involved there. Our results show
that, nevertheless, the basic properties of holographic baryons
obtained there are  robust under stringy corrections
within a few percents.
}

\preprint{
{\normalsize CERN-PH-TH-2010-062} \\
{\normalsize KIAS-P10007} \\
{\normalsize RIKEN-TH-186}\\
{\normalsize RIKEN-MP-1}\\
}

\begin{document}
\setcounter{page}{1}


\section{Introduction}
\label{sec1}

Nuclear physics is one of the oldest branches of high energy
physics, yet remains one of more difficult. Despite the fact
that we know the underlying fundamental theory, {\it i.e.} QCD,
we are still unable to predict, reliably and analytically,
behavior of nuclei or  even a single proton. The
problem is of course that one must understand the strong-coupling
regime of QCD, which by and large remains inaccessible except by
large-scale lattice simulations. Traditionally, this set
nuclear physics apart from the rest of high energy physics in many
aspects.
However, recent developments in the so-called gauge/gravity
duality began to solve certain strongly coupled field theories,
possibly including QCD or its close relatives,
allowing the two communities merging with each other.
In this paper, by using the gauge/gravity duality, we propose a new
matrix model for the dynamics of multi-baryons, whereby we compute
basic properties of holographic baryons, interaction with mesons,
and ultimately nuclear forces.

What matrix? In our matrix model, the number of baryons is represented
by the rank of the matrix. Therefore, for $k$-body baryons, it is a
$U(k)$ matrix model. If we path-integrate out the off-diagonal elements
of matrices, we are left with $k$ diagonal elements. It is these elements
which represents the positions of the $k$ baryons. In addition, there are
a pair of complex $k\times N_f$ rectangular matrices whose classical values
are related to the size of baryon. Together, they form the well-known
Atiyah-Drinfeld-Hitchin-Manin (ADHM) matrix of instantons.

Why matrix? In the large $N_c$ QCD, mesons are open strings and
light degrees of freedom, with mass $\sim {\cal O}(1)$
\cite{'tHooft:1973jz}, while baryons are
solitons with large mass $\sim {\cal O}(N_c)$ \cite{Witten:1979kh}.
If we embed the large $N_c$ gauge theory into string theory, baryons are
described by string theory solitons, {\it i.e.}, D-branes
\cite{Witten:1998xy,Gross:1998gk}. Therefore, the dynamics of
multi-baryons are described by a multi-D-brane system, which is
nothing but the $U(k)$ matrix model.
Our matrix model is along the line of
the ADHM construction \cite{Atiyah:1978ri},
which is the matrix description for multi-instantons in gauge
theory, or equivalently along the line of D0-D4 quantum
mechanics.\footnote{The relevance
of such a matrix model for baryons was previously emphasized
by one of the authors \cite{Hashimoto:2008jq,Hashimoto:2009pe}.}
In the context of QCD, one must of course
deform the description appropriately.
Before we explain these changes in detail, let us first briefly
review the holographic QCD.

The application of gauge/gravity duality
\cite{Maldacena:1997re,Gubser:1998bc,Witten:1998qj}
to large $N_c$ QCD \cite{'tHooft:1973jz}, {\it i.e.}, holographic QCD
was developed very much in recent years, especially thanks
to the D4-D8
model by Sakai and Sugimoto \cite{Sakai:2004cn}. This model starts with
large $N_c$ number of D4-branes compactified on a thermal circle
\cite{Witten:1998zw}, representing pure QCD at the low energy,
and incorporates $N_f$ species of massless quarks by intersecting $N_f$
pairs of D8- and anti-D8-branes \cite{Sakai:2004cn}. In the large $N_c$
limit, the D4-branes are replaced by their dual geometry, and flavor D8
and anti-D8-branes are connected at the IR of the dual geometry, which
is the geometrical realization of chiral
symmetry breaking. The theory on the connected D8-anti-D8 brane in this dual
geometry reproduces the low energy effective theory of light meson
sector with remarkable accuracy.

This theory has only two independent input parameters,
and are free of ambiguities which low energy chiral Lagrangians
such as Skyrme model \cite{Skyrme:1962vh} possess. Furthermore, it contains
not only the light vector meson but also infinite towers of massive
vector meson, resulting in a model with infinite number of
predictions. Most of these are irrelevant since at high energy
the theory deviates from real QCD, but nevertheless there are
many low energy processes which can be computed from this model.
This aspect of the D4-D8 model \cite{Sakai:2004cn,Sakai:2005yt} is
one of the most important points that sets the D4-D8 model
aside from the other more generic AdS-inspired models.

Baryons in this D4-D8 setup turned out to be very interesting also.
Baryons, in the large $N_c$ limit, are inherently nonperturbative
objects with their
mass scaling as $N_c$  \cite{Witten:1979kh}
which is inverse of the genus expansion coupling constant.
The holographic baryon in D4-D8 model is no exception,
and can be introduced as instantonic soliton on the flavor D8-brane.
Note that no new parameters are introduced in this step and all
observables associated with baryons are computable by the D4-D8
model.\footnote{
Employing this picture, static properties \cite{Hong:2007kx,Hata:2007mb},
interactions with mesons \cite{Hong:2007kx,Hong:2007ay}, electromagnetic
form factors \cite{Hong:2007dq,Hashimoto:2008zw},
and more recently nucleon-nucleon potential
\cite{Hashimoto:2009ys,Kim:2009sr}
have been derived again with remarkable accuracy. Although
most literatures considered nucleons, higher isospin baryons
can also be treated on equal footing
\cite{Park:2008sp,Grigoryan:2009pp}.
}

Following the original idea \cite{Witten:1998xy,Gross:1998gk},
we recall that baryons can be also
thought of as D4-branes wrapped on the compact $S^4$ that surrounds
the QCD $N_c$ D4-branes. Let us call them D4'-branes
to distinguish them from those responsible for QCD.
As such, its effective dynamics in the large $N_c$ limit involve
open strings with both ends on the D4'-branes and also those connecting the
D4'-branes and the D8-branes.
The dynamics of $k$ baryons would be $U(k)$ gauge theory, and
if we consider only the zero mode\footnote{The subtlety of
taking of a decoupling limit of the non-zero modes along $S^4$
is a long-standing problem in holographic QCD.
As a result, it is obscure why in low energy effective theories such
heavy objects like baryons are well-described. For one possible
explanation concerning supersymmetries offered by one of the authors,
see \cite{Yi:2009et}.}
along the $S^4$ on which the D4' and the D8 are wrapped
on, the theory on the D4'-branes reduces to a 0+1 dimensional
matrix model.
Motivated by this viewpoint, in this paper, we propose a new $U(k)$
matrix model for holographic baryons.

To illustrate our matrix model, let us remember a
simple D-brane bound state of the D$p$-D$(p+4)$ system and
that we have two descriptions for this system.
One is the solitonic description for the D$p$-branes as instanton solitons
in the D$(p+4)$-brane gauge theory. This description is natural
when gauge fields are weakly varying,
namely at long distance scale with $\rho \gg l_s$ where
$\rho$ is the size of the instanton soliton and $l_s$ is the string
length.
The other is the open string matrix theory viewpoint, whereby
the ADHM construction of the instanton soliton is naturally
derived \cite{Witten:1995gx,Douglas:1995bn}. This
approach is more natural
at a short distance scale, $\rho \ll l_s$. For fully
supersymmetric case, the two descriptions are equivalent for
many purposes.

Our matrix model for the $k$ D4'-branes follows the latter viewpoint
and is a D4'-D8 matrix theory compactified on a common
$S^4$.  There are several differences between
our matrix model for baryons and the
usual D$p$-D$(p+4)$-brane system.
First, in our matrix model, there is a Chern-Simons (CS)
term (supersymmetric versions were studied in
\cite{Collie:2008vc,Kim:2008kn}).
This term originates from the fact that there is a Ramond-Ramond (RR)
flux on the $S^4$ on which the D4'-branes are wrapped, and is in fact
the same type of term that allowed Witten to identify wrapped
D5-branes as baryon vertices in AdS/CFT description of maximally
supersymmetric Yang-Mills theory \cite{Witten:1998xy}. For us, this
term turns out to
play a crucial role
in dynamics of baryons in general and in the baryon-baryon interactions
in particular.
Second, since the D4'-branes are living on the D8-brane at the IR bottom
of the warped geometry,
we need to take into account the warped geometry to derive the $U(k)$
matrix model. The point is that if we put
the $k$ baryons at short distances,
the warp factor approaches almost constant values, therefore
the effects of the warped geometry is simply just rescaling of a
coupling constant plus mass terms.
This in turns implies that the supersymmetry is broken explicitly
by the dual geometry \cite{Witten:1998zw,Gibbons:1987ps}
of the QCD D4-branes. In practice, the  supersymmetry breaking would
manifest as various mass terms and potential energy in the would-be
vacuum moduli space.

One motivation for us is that such a matrix
formulation can easily accommodate a large number of baryons.
One of more grey area of nuclear physics, from the purely
theoretical viewpoint, is how one handles many-nucleon systems
such as nuclei. The fundamental theory of QCD is even less
effective there. The matrix model of ours can be written down
immediately for all $k$, and should contain in principle, not
only 2-body interactions but arbitrary $k$-body interactions
built-in from the beginning. This may open up a novel window
for dealing with finite $k$ nuclei and large $k$ physics such
as necessary for neutron star, in term of strongly-coupled QCD.

Another motivation, which is perhaps more practical for now,
is already apparent in the holographic soliton picture of baryon
\cite{Hong:2007kx,Hata:2007mb}. Remember that the holographic
size of the solitonic baryon is $\simeq 9.6/(M_{\rm KK}\sqrt\lambda)$,
where $\lambda$ is the 't Hooft coupling and $M_{\rm KK}\sim 1$ GeV
is the mass scale of the vector mesons \cite{Hong:2007kx,Hata:2007mb}.
Although this size is larger than the Compton size of the baryon
$\sim 1/(M_{\rm KK}N_c\lambda)$, obeying the usual criterion
for validity of field theory soliton picture, it is unfortunately
as small as the local string scale $l_s^{\rm eff}
\sim 1/(M_{\rm KK}\sqrt{\lambda})$. One should be dubious whether
the field theory soliton is really accurate enough for such
a small object. The supersymmetry breaking scale
is $\sim 1/M_{\rm KK}$, so the supersymmetry is approximately valid at the
very short distance such as the string scale, so one hope for
relative stability of the baryon physics in interpolating between
the string scale and the QCD meson scale \cite{Yi:2009et}. Nevertheless, we cannot
expect quantitative agreement for all observables.

Open string description of D-brane interactions is deemed to
be relevant for distance scale below string scale, so we have
a motivation and an opportunity here to reexamine
holographic baryons from a D4'-D8 (compactified on $S^4$) matrix description.
With supersymmetry broken at $M_{KK}$ and below, one should
in general expect the solitonic viewpoint and the matrix viewpoint
would disagree quantitatively;
it would be interesting to compare our matrix description with
the previous soliton description.

In section 2, we drive the matrix model by starting with D4'-D8
gauge theory compactified on $S^4$, and extracting leading order
supersymmetry-breaking effects at very short distance. The action
we find is valid at distance $l^{\rm eff}_s$ and below.
To show the effectiveness of our matrix model, we display only
two simple examples in this paper: energy functions of static
configurations for $k=1$ and $k=2$, for one and two flavor(s).

In section 3, we discuss $k=1$ and determine the
holographic ``size'' of the baryon, value of which affects
numerous observable quantities. This parallels the energy
function estimates for the solitonic baryon in spirit but differs
in detail. For two or more flavors, we find the energy function
with a different numerical coefficient, leading to a new estimate
for the holographic size which is larger by a factor of $(5/4)^{1/4}$
than the soliton result. We discuss its implications. With a singe
flavor, for which the soliton model has no computation due to
lack of finite and smooth self-dual $U(1)$ instanton solutions,
we find a smaller but still non-vanishing size.

In section 4, where $k=2$ is studied, we compute a baryon-baryon
potential at short distance for two flavors. Interestingly, integrating out the
auxiliary gauge potential in 0+1 dimension turns out to give
a universal repulsive core of the nuclear force. The core  consists of three
terms; the isospin-independent central term, isospin-dependent
central term, and isospin-dependent tensor term. The computation
here is far simpler than the soliton computation, yet gives the
same type of results, except that numerical coefficient of the
latter two turn out to be larger by a factor of $5/4$. We explain
this in simple terms based on the above $k=1$ results. We also
show that the repulsive core is universal for any baryon state
for the two-flavor case.

In section 5, we close with discussions on possible ramifications
of this new model of baryons and baryon dynamics.

\section{The matrix model}
\label{sec2}

We derive the action of the matrix quantum mechanics in this section,
which will be used for getting the baryon spectrum (section \ref{sec3}),
the nucleon-nucleon potential and the universal repulsive core
(section \ref{sec4}).

The matrix model is nothing but the low energy effective action on the
$k$ D4'-branes embedded in flavor $N_f$ D8-branes in Witten's geometry.
Those who are not interested in the string-theoretical derivation are
advised to see only the action (in section \ref{sec21}), skip the rest
and go directly to section \ref{sec3}.

\subsection{Action}
\label{sec21}

The matrix model action we derive is a $U(k)$ quantum mechanics,
\begin{eqnarray}
 S &=&
\frac{\lambda N_c M_{\rm KK}}{54 \pi}
\int\! dt \; {\rm tr}_k
\left[
(D_0 X^M)^2 -\frac23 M_{\rm KK}^2 (X^4)^2
+ D_0 \bar{w}^{\dot{\alpha}}_i D_0 w_{\dot{\alpha}i}
- \frac16  M_{\rm KK}^2 \bar{w}^{\dot{\alpha}}_i w_{\dot{\alpha}i}
\right.
\nonumber \\
&&\left.
\qquad\qquad\qquad\qquad
+ \frac{3^6 \pi^2}{4 \lambda^2 M_{\rm KK}^4}
\left(\vec{D}\right)^2
+  \vec{D}\cdot \vec{\tau}^{\;\dot{\alpha}}_{\;\;\;\dot{\beta}}
\bar{X}^{\dot{\beta}\alpha} X_{\alpha \dot{\alpha}}
+  \vec{D}\cdot \vec{\tau}^{\;\dot{\alpha}}_{\;\;\;\dot{\beta}}
\bar{w}^{\dot{\beta}}_{i} w_{\dot{\alpha}i}
\right]
\nonumber \\
&&\quad\quad +  N_c \int \! dt\; {\rm tr}_k A_0 \, .
\label{mm}
\end{eqnarray}
Here $\lambda=N_c g_{\rm QCD}^2$ is the 'tHooft coupling constant, and
$M_{\rm KK}$ is the unique dimension-ful constant.
The dynamical fields are $X^M$ and $w$, while $\vec{D}$ and $A_0$ are
auxiliary fields. All the fields are bosonic.
We claim that this matrix model describes the $k$-baryon
system, according to the holographic principle in string theory.

Our 1-dimensional
matrix model is a deformed ADHM matrix model. The ADHM matrix
model has been extensively studied in the context of D-branes in string
theory and instanton calculus (for a concise review, see
\cite{Dorey:2002ik}).
Our theory is deformed in the following two points:
\begin{itemize}
\item Addition of the CS coupling. The last term of the
      action
(\ref{mm}) is a CS term in 1 dimension.
 \item Mass deformation. A part of the dynamical fields, $X^4$ and $w$,
       are massive, in contrast to the standard ADHM matrix model.
\end{itemize}
We will describe how the ADHM matrix model and
these deformations appear in the holographic QCD, together with the detailed
derivation of the coefficients in the action. Note that in the absence
of the field $w$, the CS term and the mass term, our matrix model
looks close to the BFSS Matrix theory for M-theory \cite{Banks:1996vh}
or the IKKT matrix model \cite{Ishibashi:1996xs}, as the
integration of the auxiliary field $D$ results in a potential of a
commutator type, ${\rm tr} ([X,X]^2)$.

The symmetry of this matrix quantum mechanics is
\begin{eqnarray}
 U(k)\times SU(N_f)\times SO(3)
\end{eqnarray}
where the first $U(k)$ is a local symmetry with which the gauge field
$A_0$ is associated, and the remaining
$SU(N_f)\times SO(3)$ is a global
symmetry. $k$ refers to the number of baryons of the system, $N_f$ is
the number of flavors in QCD, and $SO(3)$ is the rotational symmetry of
our space in which the baryons live. The notation of the action
is better-understood if we embed the rotational symmetry as
\begin{eqnarray}
 SO(3) \subset SO(4) \sim SU(2)\times SU(2)
\end{eqnarray}
where the additional dimension in fact corresponds to the holographic
dimension. This $SO(4)$ is broken down to the $SO(3)$ by the mass
deformation.

We summarize the representation of the fields in the table 1. 
\begin{figure}
\begin{center}
\begin{tabular}{|c|c||c|c|c|}
\hline
 field & index & $U(k)$ & $SU(N_f)$ & $SU(2)\times SU(2)$\\\hline
\hline
$X^M(t)$ & $M=1,2,3,4$ & adj. & ${\bf 1}$& $({\bf 2}, {\bf 2})$\\\hline
$w_{\dot{\alpha}i}(t)$ & $\dot{\alpha}=1,2$; $i=1,\cdots,N_f$
& ${\bf k}$ &${\bf N_f}$ & $({\bf 1},{\bf 2})$  \\\hline
$A_0(t)$ & & adj. & ${\bf 1}$ & $({\bf 1}, {\bf 1})$ \\\hline
$D_s(t)$& $s=1,2,3$ & adj. & ${\bf 1}$& $({\bf 1},{\bf 3})$\\\hline
\end{tabular}
\caption{Fields in the matrix model.}
\end{center}
\end{figure}
The indices with respect to the $U(k)$ gauge group are implicit.
In the action, the trace is over these $U(k)$ indices.
One can think of the flavor symmetry as $U(N_f)$, while the overall
$U(1)$ part of it is identical to the overall $U(1)$ part of the gauge
symmetry $U(k)$, as seen in how they act on the bi-fundamental field $w$.

In the action, the definition of the covariant derivatives is
$D_0 X^M \equiv \partial_0 X^M -i[A_0, X^M]$,
$D_0 w \equiv \partial_0 w -i w A_0$,
$D_0 \bar w \equiv \partial_0 \bar w + i A_0 \bar w$,
and $\tau^s$ $(s=1,2,3)$ is the Pauli matrix.
The spinor indices of $X$ are defined as
$X_{\alpha\dot{\alpha}}\equiv X^M(\sigma_M)_{\alpha\dot{\alpha}}$ and
$\bar{X}^{\dot{\alpha}\alpha}
\equiv X^M(\bar{\sigma}_M)^{\dot{\alpha}\alpha}$
where $\sigma_M=(i\vec{\tau}, 1)$ and
$\bar{\sigma}_M = (-i\vec{\tau},1)$.
We follow the notation of \cite{Dorey:2002ik}.

\subsection{Derivation in gauge/gravity duality}

Our matrix model (\ref{mm}) is nothing but a low energy effective field
theory on D-branes. The D-branes of our concern are
D4'-branes wrapping $S^4$ of a background geometry given by Witten
\cite{Witten:1998zw} (the metric originally given in
\cite{Gibbons:1987ps}). This D4'-brane is called ``baryon vertex''
\cite{Witten:1998xy,Gross:1998gk} in the
gravity side of the AdS/CFT duality, which
corresponds, as the name shows, to a baryon in the field theory side.
As we are dealing with a $k$-baryon system, we place these $k$ D4'-branes
close to each other. We are going to derive the effective action of this
collection of the D4'-branes wrapping the $S^4$, via a standard technique
in string theory.

The D-brane action is affected not only by the geometry and the
background flux, but also by the presence of the probe $N_f$ D8-branes
which are responsible for quarks {\it a la} Sakai and Sugimoto
\cite{Sakai:2004cn}. We have additional strings connecting the baryon
D4'-brane and the flavor D8-brane. From the viewpoint of the baryon
D4'-brane effective field theory, this string provides the field $w$
in the bi-fundamental representation. Together with the field $X^M$ in
the adjoint representation whose diagonal eigenvalues specify the
location of the D4'-branes in the transverse directions (but longitudinal
to the D8-brane worldvolume) and thus the location of the baryons in our
real space, the matrix model action is written.

Since the D4'-branes are completely inside the worldvolume of the flavor
D8-branes, the matrix model is very close to the so-called
supersymmetric ADHM
matrix model which is nothing but the effective action of $k$
D-instantons on $N_f$ D3-branes in flat spacetime. The deformation is
due to the curved geometry and the flux, which break the supersymmetry
explicitly. Here we keep only the bosonic fields, which is enough for
computing classical quantities.

Although the D4'-branes (and the D8-branes) wrap the $S^4$, we
trivially reduce the
$S^4$ dependence (a dimensional reduction with the assumption of no
dependence along $S^4$), so that the resulting action is in one
dimension, {\it i.e.} only time direction.

The background geometry and the flux given by
Witten \cite{Witten:1998zw}
(and Gibbons and Maeda \cite{Gibbons:1987ps}) are written as
\begin{eqnarray}
&& ds^2 = (U/R)^{3/2} (\eta_{\mu\nu} dx^\mu dx^\nu + f(U)d\tau^2)
+ (R/U)^{3/2}(f(U)^{-1} dU^2 + U^2 d\Omega_4^2)\, ,
\\
&& e^\phi = g_s (U/R)^{3/4}, \quad
F_4 = dC_3 = \frac{2\pi N_c}{V_4}\epsilon_4 \, ,
\label{RR}
\end{eqnarray}
where $f(U)\equiv 1-U_{\rm KK}^3/U^3$, and
$R^3 \equiv \pi g_s N_c l_s^3$.
$V_4 \equiv 8\pi^2/3$ is the volume of the $S^4$,
and $\epsilon_4$ is the volume form on it.\footnote{
We use the standard normalization for the forms,
$C_3 = (1/3!)\; C_{ijk}dx^i\wedge dx^j\wedge dx^k$ and
$F_4 = (1/3!)\; \partial_\alpha
C_{ijk}dx^\alpha\wedge dx^i\wedge dx^j\wedge dx^k$.}
The $\tau$ direction is compactified with the period
$\tau \sim \tau + 2\pi/M_{\rm KK}$ where $M_{\rm KK}\equiv
(3/2)U_{\rm KK}^{1/2} R^{-3/2}$,
so that the geometry is everywhere smooth.
The relations to the QCD variables are
\begin{eqnarray}
 R^3 = \frac12 \frac{g_{\rm YM}^2 N_c}{M_{\rm KK}}l_s^2\, ,\quad
 U_{\rm KK} = \frac29 g_{\rm YM}^2 N_c M_{\rm KK} l_s^2\, ,\quad
 g_s = \frac{1}{2\pi} \frac{g_{\rm YM}^2}{M_{\rm KK}} \frac{1}{l_s}\, .
\end{eqnarray}
Convenient coordinates used in \cite{Sakai:2004cn,Sakai:2005yt} are
\begin{eqnarray}
&& U^3 = U_{\rm KK}^3 + U_{\rm KK}r^2\, ,
\quad \theta \equiv \frac{3}{2}
\frac{U_{\rm KK}^{1/2}}{R^{3/2}}\tau\, ,
\quad y+ iz \equiv r e^{i\theta}\, .
\end{eqnarray}
The flavor D8-branes are located at $y=0$.

Below we derive the action, by looking at, first,
the background RR flux,
and second,
the effect of the background geometry.

\vspace{3mm}

\noindent
\underline{Chern-Simons term}

The important term is the last term of the matrix model (\ref{mm}),
which is a CS term in 0+1 dimension.
Using the background RR flux (\ref{RR})\footnote{
We are using the normalization of the RR field in which
the RR charge is measured in units of $2\pi$, see Appendix A of
\cite{Sakai:2004cn}.}
we can compute the
CS term on the D4'-brane as
\begin{eqnarray}
 S &=& \frac{1}{2\pi} \frac{1}{2 \cdot 3!}
\int \! d^5\xi \; {\rm tr}
\; \epsilon^{\mu_1 \mu_2 \mu_3 \alpha \beta}
C_{\mu_1\mu_2\mu_3}F_{\alpha\beta}
 =
\frac{1}{2\pi}\! \int \! dt \;
{\rm tr}A_0 \int \!F_4
\nonumber \\
 &=&
N_c\! \int \! dt \; {\rm tr}A_0 \, .
\end{eqnarray}
Note that the overall factor $N_c$ shows that the D4'-brane should be
supplied with $N_c$ fundamental strings (the end point of the D4'-D8
string serves as an electric charge on the D4'-brane), which means that
the D4'-brane is indeed a baryon \cite{Witten:1998xy,Gross:1998gk}.\footnote{
This Chern-Simons term is also important in producing the correct statistics
of baryons, which should be either fermionic or bosonic depending on whether
$N_c$ is odd or even. For detail of nucleon statistics in our matrix model, see Ref.~\cite{Hashimoto:2010rb}.}

Interestingly, this 0+1-dimensional CS term was used in
\cite{Collie:2008vc} and \cite{Kim:2008kn} for the ADHM matrix model for
supersymmetric dyonic instantons with a CS term in 5 dimensions. There, the CS term was
argued for heuristically (\cite{Collie:2008vc} studied fermions and
anomalies to reach the 0+1-dimensional CS term). Here we have derived
the 0+1-dimensional CS term from string theory.

\vspace{3mm}

\noindent
\underline{Mass terms and overall normalization}

The Dirac-Born-Infeld part of the action for a single D4'-brane is
\begin{eqnarray}
 S = -T_{\rm D4}\int\! d^5\xi\; e^{-\phi}\;
\sqrt{- \det(G_{MN}+ 2\pi\alpha' F_{MN})} \, .
\label{DBI}
\end{eqnarray}
We consider a D4'-brane situated at $y=0$ which wraps the $S^4$.
Then
\begin{eqnarray}
 S = -\frac{T_{\rm D4}}{g_s} \int \! dt\;
V_4 \left((R/U)^{3/2} U^2\right)^2
(U/R)^{-3/4}\sqrt{-G_{00}}\,
\end{eqnarray}
with $V_4\equiv 8\pi^2/3$ the volume of a unit four-sphere,
where the induced metric is
\begin{eqnarray}
 G_{00} = - \left(\frac{U}{R}\right)^{3/2}
\left(1-(\partial_0 X^i)^2\right) + \frac49
\left(\frac{R}{U}\right)^{3/2}
\frac{U_{\rm KK}}{U} (\partial_0 Z)^2 \, .
\end{eqnarray}
The index of $X^i$ runs for our 3-dimensional space, $i=1,2,3$.
So, we obtain
\begin{eqnarray}\label{detailedS}
 S = -\frac{T_{\rm D4}}{g_s} \int\! dt \;
\frac{8\pi^2}{3}R^3 U
\sqrt{
1-(\partial_0 X^i)^2 - \frac49
\frac{R^3 U_{\rm KK}}{U^4}
 (\partial_0 Z)^2
} \, .
\end{eqnarray}
with $T_{Dp}=2\pi/(2\pi\sqrt{\alpha'})^{p+1}$.

We expand this for small $Z$ and small $X$. Using the expansion
$U = U_{\rm KK} (1 + (1/3)U_{\rm KK}^{-2} Z^2 + {\cal O}(Z^4))$,
and a redefinition
\begin{eqnarray}
X^4
\equiv \frac23 \left(\frac{R}{U_{\rm KK}}\right)^{3/2}Z
\, ,
\end{eqnarray}
we obtain a quadratic Lagrangian
\begin{eqnarray}\label{XZ}
 S = \frac{\lambda N_c M_{\rm KK}}{27\pi}
\int \! dt\left[-1+
\frac12 (\p_0 X^i)^2 +\frac12
(\p_0 X^4)^2
- \frac13 M_{\rm KK}^2 (X^4)^2 
\right]\, .
\end{eqnarray}
In this way, the overall normalization of the matrix model
action,\footnote{
Note that higher order terms in
$X^4$
are not suppressed by
$1/\lambda$. In fact, the leading correction to the mass term is
\begin{eqnarray}
 + \frac13 M_{\rm KK}^2  (X^4)^2  
- \frac19 M_{\rm KK}^4 (X^4)^4 
+ \cdots
\end{eqnarray}
In the following, we assume that the magnitude of
$X^4$
is
small so that we can ignore the higher corrections. In fact, as we
consider the wave function of this
$X^4$
after the quantization around the vacuum of the matrix model, we obtain
a Gaussian wave function with the width suppressed by $1/N_c$ which is
quite small, and this approximation is valid.
}
as well as the mass term for the field
$X^4$,
are provided in the matrix model (\ref{mm}). This mass term originates
entirely from the expansion of $U$ sitting in front of the square-root in
(\ref{detailedS}).

This leaves the mass term for $w$.
We assume that it is
\begin{eqnarray}
 - \frac{\lambda N_c M_{\rm KK}}{54\pi}\int \! dt \;
\frac16  M_{\rm KK}^2 \bar{w}^{\dot{\alpha}}_i w_{\dot{\alpha}i}
\label{wmassterm}
\end{eqnarray}
to be added to (\ref{XZ}), which completes the quadratic part of the
matrix action in (\ref{mm}). This (\ref{wmassterm}) is an educated guess
via a comparison with the soliton picture (though in principle one can
derive this by computing string scattering amplitudes).
In fact, as we will see in
the next section, the
$X^4$ 
mass term in (\ref{XZ}) coincides
with the soliton picture, and the following argument suggests that
$w$ mass also coincides.

Recall that
$(X^i, X^4, w)$
form the
ADHM data for instantons. When the instanton size is very small compared
to the supersymmetry-breaking scale $1/M_{KK}$, the background geometry
is effectively flat with approximate supersymmetry. The D4'-D8 system
will inherit supersymmetry, and thus the data
$(X^i,X^4,w)$
is equivalent
to the supersymmetric instanton on the D8-branes; At this zero-th order, the
instanton picture of the D4'-brane on the D8-branes is still valid. Note that this
is not yet the baryon but merely a purely magnetic instanton solution.
Then, we turn to the effect of supersymmetry breaking at $M_{KK}$ scale, and
evaluate the potential energy of the instanton at the first order.
This is in fact one way of obtaining the mass term for
$X^4$
in (\ref{mm}).
Similarly, using the fact that $|w|^2/2$ is the size squared for a single
instanton, we find (\ref{wmassterm}) for the matrix model.\footnote{
One may recall that the soliton representation of the holographic baryon
has a Coulombic potential as well \cite{Hong:2007kx,Hata:2007mb}.
This is not apparent yet in the matrix model but will arise by integrating
out the gauge potential $A_0$ of D4'. The crucial difference from the mass
terms is that the Coulombic energy arises from a quadrature of an electric
field (in the soliton picture) which is itself a first order deviation.
We will see later that, perhaps because of this, the Coulombic energy differs
quantitatively in the two pictures.
}


\vspace{3mm}

\noindent
\underline{Commutator term}

Finally, let us compute the coefficient in front of $\vec{D}^2$ which is
related to the coefficient in front of the famous commutator term
$[X,X]^2$ in D-brane quantum mechanics.

Let us expand the generic D$p$-brane action (\ref{DBI})
to the quadratic order,
\begin{eqnarray}
 S = -T_{{\rm D}p} \int\! d^{p+1}\xi \; e^{-\phi}
\sqrt{- \det G_{MN}} \frac14 (2\pi\alpha')^2
F_{MN}F_{PQ} G^{MP} G^{NQ} \, .
\end{eqnarray}
We now make a dimensional reduction to get the commutator term from the
YM kinetic action. The relevant formula for the dimensional reduction is
$2\pi\alpha' A_M = X^N G_{MN}$
for diagonal metrics. Then, the action is
\begin{eqnarray}
 L &\;\propto \;&
2 G^{00} G_{ij} D_0 X^i D_0 X^j
+2 G^{00} G_{zz} D_0 Z D_0 Z
\nonumber \\
& &
-[X^i, X^j][X^k, X^l]G_{ik}G_{jl} \frac{1}{(2\pi\alpha')^2}
-2[X^i, Z][X^j, Z]G_{ij}G_{zz} \frac{1}{(2\pi\alpha')^2}
\nonumber \\
& \propto &
(D_0 X^i)^2
+ (D_0 X^4)^2
+ \frac12 \frac{1}{(2\pi\alpha')^2}
\left(
[X^i,X^j]^2 + 2[X^i, X^4]^2
\right) \left(\frac{U_{\rm KK}}{R}\right)^3
\nonumber \\
& = &
(D_0 X^M)^2
+ \frac{2}{3^6\pi^2}\lambda^2 M_{\rm KK}^4
[X^M,X^N]^2 
\, .
\label{com}
\end{eqnarray}

On the other hand, the commutator term can be written by using the
auxiliary field $\vec{D}$. If we start from the action of
\cite{Dorey:2002ik}\footnote{
In \cite{Dorey:2002ik} the field $\vec{D}$ is anti-Hermitian, while our
$\vec{D}$ is defined to be Hermitian.
}
\begin{eqnarray}
 S = c \int \! dt \; {\rm tr}
\left[
2 (2\pi\alpha')^2(\vec{D})^2 +  \vec{D}\cdot
\vec{\tau}^{\;\dot{\alpha}}_{\;\;\;\dot{\beta}}
\bar{a'}^{\dot{\beta}\alpha}a'_{\alpha \dot{\alpha}}
\right]
\end{eqnarray}
then by integrating out the field $\vec{D}$ we obtain
\begin{eqnarray}
 S = c \int\! dt \; {\rm tr}
\left[\frac{1}{16 \pi^2 \alpha'^2}
[a'_m, a'_n]^2
\right] \, .
\end{eqnarray}
Comparing this with the normalization we obtained in (\ref{com}),
we obtain the expression for the commutator term of our matrix model as
\begin{eqnarray}
\frac{\lambda N_cM_{\rm KK}}{54 \pi}
\int\! dt \; {\rm tr}_k
\left[
 \frac{3^6 \pi^2}{4 \lambda^2 M_{\rm KK}^4}
\left(\vec{D}\right)^2
+  \vec{D}\cdot \vec{\tau}^{\;\dot{\alpha}}_{\;\;\;\dot{\beta}}
\bar{X}^{\dot{\beta}\alpha} X_{\alpha \dot{\alpha}}
\right] \, .
\end{eqnarray}
The $w$ term coupled to $\vec{D}$ is written down in \cite{Dorey:2002ik}
and we can just use it with the same normalization as the $\vec{D}XX$
term.

\section{Single baryon}
\label{sec3}

In this section, we study the $k=1$ case, {\it i.e.} a single baryon.
Our quantum mechanics directly gives a spectrum of the baryon.

First, we evaluate the Hamiltonian of the quantum mechanics with
$k=1$. Then we analyze the vacuum of the
system for $N_f=1$ and $N_f=2$ respectively.
For  $N_f=2$ system, we calculate the baryon
spectrum. The computation of the spectrum, as well as the quantization
procedure, closely follow the soliton approach of
\cite{Hong:2007kx,Hata:2007mb}, although the derivation of the Hamiltonian
 is different. Finally we discuss meson couplings.

\subsection{Hamiltonian}

Let us compute the Hamiltonian for a single baryon $k=1$, with generic
$N_f$.

First we explicitly integrate out auxiliary fields $\vec{D}$ and $A_0$.
As for the terms including the field $\vec{D}$,
since for $k=1$ the field
$X$ is now not a matrix but a number, all the $X$ couplings
drop off, and we obtain
\begin{eqnarray}
 S_{\vec{D}} &=&\frac{\lambda N_c M_{\rm KK}}{54 \pi}
\int \! dt \;
\frac{-\lambda^2 M_{\rm KK}^4}{3^6 \pi^2}
\left[
\sum
\left( \vec{\tau}^{\;\dot{\alpha}}_{\;\;\;\dot{\beta}}
\bar{w}^{\dot{\beta}}_{i} w_{\dot{\alpha}i} \right)^2
\right]
\nonumber \\
&=& \frac{\lambda N_c M_{\rm KK}}{54 \pi}
\int \! dt \;
\frac{-\lambda^2 M_{\rm KK}^4}{3^6 \pi^2}
\left[
4 w_1^i (w_2^i)^* w_2^j (w_1^j)^*
\right.
\nonumber \\
&&
\left.
\hspace{30mm}
+ (w_1^i (w_1^i)^*)^2 + (w_2^i (w_2^i)^*)^2
-2 w_1^i (w_1^i)^*w_2^j (w_2^j)^*
\right] \,,
\end{eqnarray}
where, the first sum is over three Pauli matrices and we omit the dots
in the dotted spinor, $\dot{\alpha}=1,2$.
$S_{\vec{D}}$ gives a so-called ADHM potential.
Minimization of the ADHM potential is equivalent to the ADHM constraint,
which should be solved for construction of instantons in the ADHM
formalism. Note that since Lagrangian is Hermitian,
$\vec{\tau}^{\;\dot{\alpha}}_{\;\;\;\dot{\beta}}
\bar{X}^{\dot{\beta}\alpha} X_{\alpha \dot{\alpha}}$
and
$\vec{\tau}^{\;\dot{\alpha}}_{\;\;\;\dot{\beta}}
\bar{w}^{\dot{\beta}}_{i} w_{\dot{\alpha}i}
$
are real.

Since our theory is 0+1-dimensional, the gauge field $A_0$ is an auxiliary
field, and we integrate it out explicitly.
The terms including $A_0$ in the matrix model action is
\begin{eqnarray}
S_{A_0} &=&
\frac{\lambda N_c M_{\rm KK}}{54 \pi}  \!
\int \! dt
\left[
\partial_0 \bar{w}^{\dot{\alpha}} (-i)w_{\dot{\alpha}}A_0
+ i A_0 \bar{w}^{\dot{\alpha}} \partial_0w_{\dot{\alpha}}
+ (A_0)^2 \bar{w}^{\dot{\alpha}}w_{\dot{\alpha}}
+ \frac{54\pi}{\lambda M_{\rm KK}} A_0
\right] \hspace{5mm}
\label{Azeropart}
\end{eqnarray}
So, the equation of motion for this $A_0$, in other words, the Gauss law
constraint, is
\begin{eqnarray}
\frac{54\pi}{\lambda M_{\rm KK}} + i
\left( \bar{w}^{\dot{\alpha}}_i \partial_0w_{\dot{\alpha}}^i
- \partial_0 \bar{w}^{\dot{\alpha}}_i w_{\dot{\alpha}}^i \right)
+ 2 \bar{w}^{\dot{\alpha}}_i w_{\dot{\alpha}}^i A_0 =0 \, ,
\label{generalconstr}
\end{eqnarray}
Then after path-integration over $A_0$, we obtain
\begin{eqnarray}
S_{A_0} =
\frac{\lambda N_c}{54 \pi} M_{\rm KK}
\int \! d\xi^0
\left[  - \frac{1}{4 \bar{w}^{\dot{\alpha}}_i w^i_{\dot{\alpha}}}
\left(
\frac{54\pi}{\lambda M_{\rm KK}} + i
\left( \bar{w}^{\dot{\alpha}}_i \partial_0w_{\dot{\alpha}}^i
- \partial_0 \bar{w}^{\dot{\alpha}}_i w_{\dot{\alpha}}^i \right)
\right)^2
\right]\,.
\end{eqnarray}

Using the definition of the momentum conjugate to the field $w$
\begin{eqnarray}
 P^{\dot{\alpha}}_i \equiv
\frac{\p S}{\p \dot{w}_{\dot\alpha}}^i
= \frac{\lambda N_c M_{\rm KK}}{54\pi}
\left[
\partial_0 \bar{w}^{\dot\alpha}_{i}
\!-\!\frac{2}{4 \bar{w}^{\dot{\gamma}}_j w_{\dot\gamma}^j}
\left(
\frac{54\pi}{\lambda M_{\rm KK}} \!+\!
i(\bar{w}^{\dot\beta}_k\p_0 w_{\dot\beta}^k
\!-\!\p_0 \bar{w}^{\dot\beta}_k w_{\dot\beta}^k)
\right)i \bar{w}^{\dot\alpha}_i
\right]\, ,
\end{eqnarray}
we obtain the Hamiltonian
\begin{eqnarray}
 H&  \equiv &
P^{\dot\alpha}_i \dot{w}^{\dot\alpha}_i
+ \bar{P}_{\dot\alpha}^i \dot{\bar{w}}_{\dot\alpha}^i - L
\nonumber \\
& = &
\frac{\lambda N_cM_{\rm KK}}{54\pi}
\left[
\p_0 \bar{w}^{\dot{\alpha}}_i \p_0 w^i_{\dot{\alpha}}
+ \frac16 M_{\rm KK}^2
\bar{w}^{\dot{\alpha}}_i w^i_{\dot{\alpha}}
\right.
\nonumber \\
&&
+\frac{\lambda^2 M_{\rm KK}^4}{3^6 \pi^2}
\left[
4 w_1^i (w_2^i)^* w_2^j (w_1^j)^*
+ (w_1^i (w_1^i)^*)^2 + (w_2^i (w_2^i)^*)^2
-2 w_1^i (w_1^i)^*w_2^j (w_2^j)^*
\right]
\nonumber \\
&&
\hspace{10mm}
\left.
+\frac{1}{4 \bar{w}^{\dot{\alpha}}_i w^i_{\dot{\alpha}}}
\left(
\left( \frac{54\pi}{\lambda M_{\rm KK}} \right)^2 +  \left(
\bar{w}^{\dot{\alpha}}_i \partial_0w_{\dot{\alpha}}^i
-\partial_0 \bar{w}^{\dot{\alpha}}_i w_{\dot{\alpha}}^i \right)^2
\right)
\right] \, .
\label{generalhamiltonian}
\end{eqnarray}

\subsection{Single flavor}
\label{sec2.2}

Let us minimize the Hamiltonian to find a vacuum of the $k=1$ system.
We consider first the case of the single flavor, $N_f=1$.
We put the following ansatz,\footnote{
A more general ansatz is,
\begin{eqnarray}
 w_{\dot\alpha=1} = \rho_1 e^{i (v_1 t+s_1)}\, , \quad
 w_{\dot\alpha=2} = \rho_2 e^{i (v_2 t+s_2)}\, ,
\end{eqnarray}
where $\rho$'s, $v$'s and $s$'s are real constants.
However, this results in the vacuum which is the
same as what we will find below in this section.
In fact, if we minimize the Hamiltonian $H$ with respect to
$v_1$ and $v_2$, we find $v_1=v_2$,
and the Hamiltonian $H$ is independent of $v$. This is interpreted as a
manifestation of the gauge invariance. As a result, we can always choose
a gauge
$v_1=v_2=0$.
Note that if one choose another gauge, for example, $A_0=0$, then the
Gauss law
constraint (\ref{generalconstr}) forces $w$ to have a time-dependence,
{\it i.e.} $v_1 = v_2 \neq 0$ to satisfy the Gauss law constraint
(\ref{generalconstr}).
Simplified ansatz (\ref{simpleansatz})
is not consistent with a generic gauge choice for $A_0$.
This is in good contrast to the situation of the ADHM vacuum for
supersymmetric Yang-Mills-Chern-Simons instanton studied in
\cite{Collie:2008vc,Kim:2008kn}.}
\begin{eqnarray}
 w_{\dot\alpha=1} = \rho_1\, , \quad
 w_{\dot\alpha=2} = \rho_2\, ,
\label{simpleansatz}
\end{eqnarray}
where $\rho$'s  are real constants.
Then, the Hamiltonian is
\begin{eqnarray}
 H =
\frac{\lambda N_cM_{\rm KK}}{54\pi}
\left[
\frac{1}{2}  \left( \frac{27\pi}{\lambda M_{\rm KK}}\right)^2 \rho^{-2}
+ \frac13 M_{\rm KK}^2 \rho^2 + \frac{4\lambda^2 M_{\rm KK}^4}{3^6 \pi}
\rho^4
\right]
\label{hamb1}
\end{eqnarray}
where we have defined
$2\rho^2 \equiv \rho_1^2 + \rho_2^2$.

Each term in the Hamiltonian (\ref{hamb1}) has a physical meaning.
\begin{itemize}
 \item
The first term $\propto \rho^{-2}$ is induced by the CS term and the $A_0$
       path-integration, and it can be interpreted as a
self-repulsion of the dyonic instanton in the soliton picture.
As described in \cite{Hong:2007kx,Hata:2007mb},
the instanton has an electric charge, so the
       self-energy
should be lowered by expanding the size of the instanton, thus resulting
       in a negative power in $\rho$.
\item
The second term is from the mass term of our matrix model, thus comes
     from the curved spacetime of the background. In terms of the
     instanton, the location of the instanton along the direction $z$
affects the total mass of the instanton, due to the curved geometry.
\item
The third term $\propto \rho^4$ is from a path-integration over
      auxiliary $\vec{D}$ fields
which for example, yielded the commutator square term in the
      matrix model action.
So this corresponds to the ADHM potential term.
      For the single flavor, there is no U(1)
      instanton except for the small instanton singularity in flat
      space, and this term ensures it, in the absence of the dyonic
      coupling and the curved geometry.
\end{itemize}

The Hamiltonian is minimized at a nonzero $\rho$, but the minimization
problem is a non-linear equation. With a help of the fact that we are
working in the large $\lambda$ limit, we reduce the problem to a linear
one. We put
\begin{eqnarray}
 \rho = x \lambda^\alpha M_{\rm KK}^{-1}
\end{eqnarray}
where $x$ is a constant coefficient, and $\alpha$ is a constant power.
Then, each term in the Hamiltonian (\ref{hamb1}) scales for large
$\lambda$ as
\begin{eqnarray}
\frac12 \left(
\frac{27\pi}{\lambda M_{\rm KK}}\right)^2 \rho^{-2}
\sim \lambda^{-2-2\alpha}\, , \quad
\frac13 M_{\rm KK}^2 \rho^2
\sim \lambda^{2\alpha}\, ,
\quad
\frac{4\lambda^2 M_{\rm KK}^4}{3^6 \pi}
\rho^4
\sim \lambda^{4\alpha+2} \, .
\end{eqnarray}
In minimizing the Hamiltonian,
the first term, which has a negative power of $\rho$, should be
balanced with either the second or the third term which has a positive
power of $\rho$.
If
it is with
the second term, then
from the above
$\lambda$-scaling we need to have $-2-2\alpha = 2\alpha$, thus
$\alpha = -1/2$. However this value means that the third term has a
power larger than the first and the second terms, so leading to an
inconsistency.
Therefore, we conclude that the minimized Hamiltonian is
dominated by a cancellation of the first and the third term.
This leads to $-2-2\alpha = 4 \alpha +2$ which is solved as $\alpha =
-2/3$ at which indeed the second term gives a smaller contribution, which
is consistent. For this reason, we can safely ignore the second term
at the large $\lambda$, so
\begin{eqnarray}
 H \simeq
\frac{\lambda N_cM_{\rm KK}}{54\pi}
\left[\frac12
 \left(
\frac{27\pi}{\lambda M_{\rm KK}}\right)^2 \rho^{-2}
 + \frac{4\lambda^2 M_{\rm KK}^4}{3^6 \pi}
\rho^4
\right]\, .
\label{hamb1-2}
\end{eqnarray}
Ignoring the second term means that the ADHM-like potential (the third
term) is much larger than the curvature scale of the background
spacetime. This is natural, since the ADHM-like potential has the scale
of the string length, in D-brane effective actions. We will see in the next that for
the two flavor case the ADHM-like potential can vanish so
that finally the Hamiltonian is minimized by the cancellation of the
first and the second terms.

The value of $x$ minimizing this Hamiltonian is computed, as
\begin{eqnarray}
 \rho = 2^{-2/3}9\sqrt{\pi} \lambda^{-2/3} M^{-1}_{\rm KK} \, .
\end{eqnarray}
The minimized value of the Hamiltonian is
\begin{eqnarray}
 H_{\rm min} = 2^{-5/3}\lambda^{1/3} N_c M_{\rm KK} \, .
\end{eqnarray}

\subsection{Two flavors} \label{2f}

Let us consider the more realistic two-flavor case.
First, to eliminate the
contribution from the ADHM potential term (the third term), we need to
satisfy the ADHM constraints,
$\vec{\tau}^{\;\dot{\alpha}}_{\;\;\;\dot{\beta}}
\bar{w}^{\dot{\beta}}_{i} w_{\dot{\alpha}i} = 0$ for all
Pauli matrix directions, or equivalently,
\begin{eqnarray}
\sum_{i=1}^{N_f} w_{\dot \alpha =1}^i (w_{\dot \alpha =2}^i)^*
= \sum_{i=1}^{N_f} w_{\dot \alpha =2}^i (w_{\dot \alpha =1}^i)^* =0 \, ,
 \quad \sum_{i=1}^{N_f} |w_{\dot \alpha = 1}^i|^2 =\sum_{i=1}^{N_f}
|w_{\dot \alpha = 2}^i|^2\, .
\end{eqnarray}
Once this condition is met, the ADHM potential disappears, and
the total energy is lowered drastically as the power in $\lambda$
changes. This can be achieved by the following generic choice
\begin{eqnarray}
 w^{i=1}_{\dot \alpha} = \left(
 \begin{array}{c}
\rho
\\ 0
 \end{array}
\right)_{\dot \alpha}, \quad
 w_{\dot \alpha}^{i=2} = \left(
 \begin{array}{c}
0 \\\rho
 \end{array}
\right)_{\dot \alpha} \,.
\end{eqnarray}
Note that this is a generic solution
minimizing the ADHM potential,
since the condition is invariant under the $U(2)$
global transformation on the spinor index $\dot{\alpha}$ and the
$U(2)$ flavor symmetry,
\begin{eqnarray}
w_{\dot \alpha}^i \to U_{\dot \alpha}^{\;\;\;\dot \beta}
w_{\dot \beta}^j [U^{\dagger}_{\rm f}]_j^{\;\;\; i} \, .
\label{alphaglobal}
\end{eqnarray}

Then the Hamiltonian, after we include
$X^4$-dependence as well, is
\begin{eqnarray}
 H =
\frac{\lambda N_c M_{\rm KK}}{54\pi}\left[
\left(\frac{27\pi}{\lambda M_{\rm KK}}\right)^2
\frac{1}{2 \rho^2}
+ \frac13 M_{\rm KK}^2 \rho^2+ \frac23 M_{\rm KK}^2 (X^4)^2
\right]\, .
\label{hamilours}
\end{eqnarray}
This is minimized at
\begin{eqnarray}
 \rho = 2^{-1/4}3^{7/4} \sqrt{\pi} \lambda^{-1/2} M_{\rm KK}^{-1}\, . \label{size}
\label{minrho}
\end{eqnarray}
The minimized value of the Hamiltonian is
\begin{eqnarray}
 H_{\rm min} = 6^{-1/2} N_c M_{\rm KK}\, .
\end{eqnarray}
This is independent of $\lambda$, thus in the large $\lambda$ limit,
we see that  $H_{\rm min}$ for the two flavor case is far smaller than
that of the single-flavor case. Recall that the classical mass of
the holographic baryon is $\lambda N_c M_{\rm KK}/27\pi+H_{\rm  min}$
where the first term comes from the constant part of (\ref{XZ}).


Here, the variable $\rho$ is nothing but the instanton size in the
soliton approach, since we have chosen a correct normalization
for this $\rho$, {\it a la} ADHM formalism in the flat spacetime.
Let us compare our Hamiltonian with the one obtained in the soliton
approach, \cite{Hong:2007kx,Hata:2007mb}, where
a potential for moduli of the single instanton solution was computed.
The Hamiltonian \cite{Hong:2007kx,Hata:2007mb} for the instanton
size modulus $\rho$ and the instanton location
$Z$ along the $x^4$ direction is
\begin{eqnarray}
 H_{\rm soliton} = 
\frac{\lambda N_c M_{\rm KK}}{54\pi}\left[
\frac{2\cdot 3^6 \pi^2}{5\left({\lambda M_{\rm KK}}\right)^2}
\frac{1}{\rho^2}
+ \frac13 M_{\rm KK}^2 \rho^2
+ \frac23 M_{\rm KK}^2 (X^4)^2 
\right] \, ,
\label{hssy}
\end{eqnarray}
again without the rest mass term $\lambda N_c M_{\rm KK}/{27\pi}$.
We first note that
the quadratic terms in $X^4$
and in $\rho$ coincide with ours.
The coincidence of the
$X^4$
mass term is nontrivial, while the one
for $\rho$ is not accidental, since we have computed the $\rho$
mass term in (\ref{mm})
by resorting to the soliton picture in the supersymmetric limit. In some
sense, they are first order terms whose evaluation used the zero-th order
solution. The real comparison is with the first terms which are
proportional to $1/\lambda^2$ (times the naive baryon mass computed
from $D4'$ tension); the two are structurally identical but the matrix model
result is larger than the soliton result by a factor of $5/4$.

In the soliton picture, this term arises as the five-dimensional
Coulomb energy associated with $U(1)$ baryon charge; the latter is, due to
the holographic map, a gauge charge of real gauge field on $D8$. In the matrix
model, it comes from integrating out the (non-dynamical) $A_0$ gauge field on $D4'$.
Either way, it comes from a quadrature of an excitation field of order $1/\lambda$,
which suggests that, in the supersymmetric limit of very small instanton,
the Coulombic energy captures the deviation from the zero-th
order ``instanton = ADHM" configuration more effectively. We suspect
that this explains the numerical difference.
At any rate,  $5/4$ is fairly close to $1$,
implying that the difference between the two approaches is relatively
minor.
Since the two such models in general live in two vastly different validity
regions, respectively, one could have expected a larger difference. It is
the presence of approximate supersymmetry in distance scale from $1/M_{KK}$
down to $l^{\rm eff}_s$ and below that give us this relative stability.
The baryon physics of one does not deviate a lot from the other.

\subsection{Quantization}

With this result on the matrix model vacuum in mind,
we can now quantize the small fluctuations
of the $k=1$ $N_f=2$ matrix model. This spectrum should
correspond to the baryon spectrum.

As we have seen, what is different from \cite{Hata:2007mb} is just the
coefficient in front of the $1/\rho^2$ term in the Hamiltonian. So,
we can just track the difference in the computations of
\cite{Hata:2007mb} and find the spectrum of our matrix model. This
difference reflects in the constant $Q$ in \cite{Hata:2007mb},
which is now multiplied by 5/4 in our case.
Then the mass formula for the baryon excitation is
\begin{eqnarray}
 M = M_0 + \sqrt{\frac{(l+1)^2}{6} +  \frac{N_c^2}{6}}
+ \frac{2(n_\rho +
  n_Z)+2}{\sqrt{6}}\, .
\end{eqnarray}
Here $n_\rho,n_Z = 0,1,2,\cdots$ and $l=I/2=J/2$ with spin $J$ and
isospin $I$.
The difference from \cite{Hata:2007mb} is just the coefficient of
$N_c^2$ in this expression. $M_0$ is the mass of the D4'-brane which is
equal to the first term in (\ref{XZ}), $\lambda N_c M_{\rm KK}/27\pi$.

This formula in particular means that we can obtain
\begin{eqnarray}
 M_{l=3}-M_{l=1} = 0.5693 M_{\rm KK}\, .
\end{eqnarray}
If we use $M_{\rm KK}=945$[MeV] which is fit by the $\rho$ meson
mass, we obtain
\begin{eqnarray}
 M_{\Delta}-M_{N/P} = 540 {\rm [MeV]}\, .
\end{eqnarray}
This is larger than the experimental value $292$ [MeV]. The situation is
similar to the soliton approach
\cite{Hata:2007mb} which gave the value $569$ [MeV].

\subsection{Meson couplings}\label{mesonsection}

As we noted toward the end of subsection~(\ref{2f}), the energetics
of $\rho$ here differ quantitatively from the previous estimate
based on the instanton soliton picture in D8-brane gauge theory.
One consequence is that the classical value of $\rho$, which is
the holographic size of the nucleon, is slightly larger that its
previous estimates in Refs.~\cite{Hong:2007kx,Hata:2007mb}.
Denoting the latter by $\rho_{\rm soliton}$, we found
\begin{equation}
\rho^2=\sqrt{\frac{5}{4}}\,\rho_{\rm soliton}^2\,.
\label{classicalrhovalue}
\end{equation}
This new estimate modified the baryon spectra as we just saw
above, but it should also affect couplings to mesons.

This can be seen most clearly from the effective action approach
\cite{Hong:2007kx,Hong:2007ay}, where a tree-level effective
action in the five-dimensional bulk captures nucleon ${\cal N}$
coupled to mesons
\begin{eqnarray}\label{5Deff}
&&\int \left[-i\bar{\cal N}\gamma^m (\partial_m-i{\cal A}^{U(2)}_m) {\cal N}
-i m_{\cal N}\bar{\cal N}{\cal N} +{2\pi^2\rho_{\rm soliton}^2\over
3e^2}\bar{\cal N}\gamma^{mn} F^{SU(2)}_{mn}{\cal N} \right]
\,.
\end{eqnarray}
$m_{\cal N}$ and $e^2$ are known functions of the holographic
$Z$-coordinate, determined by the dual geometry. Actual
four-dimensional nucleon is the lowest lying mode of five-dimensional
Dirac field ${\cal N}$ (here, denoted
by the same symbol in abuse of notation),
while the infinite tower of mesons are embedded
into the five-dimensional flavor gauge field ${\cal A}^{U(2)}=A^{U(1)}+A^{SU(2)}$
as in Ref.~\cite{Sakai:2004cn}. Note that
the size parameter $\rho^2_{\rm soliton}$ appears explicitly only once, in the last term, so the
holographic size of the nucleon will affect couplings to $SU(2)$
iso-triplet meson coupling with specific chiral or tensor structures only.

Previous estimates of the meson-nucleon-nucleon couplings were based on
the instanton soliton viewpoint and may not be completely
compatible with our new matrix model. Nevertheless, we note
that these couplings were read-off entirely from the long distance
gauge field configuration associated with the baryon \cite{Hong:2007kx,Hong:2007ay},
{\it which
has to be the case since the Compton wavelength of mesons at
$\sim 1/M_{\rm KK}$ are much larger than the size of the soliton core
$\rho_{\rm soliton}\sim 1/M_{KK}\sqrt{\lambda}$.}
This suggests that the derivations of meson-baryon coupling are generally
safe from short-distance physics. The one place where this reasoning
can go wrong is the size of the soliton itself, whose estimate relies
heavily on short distance physics and which in turn affect
the long distance magnetic field of the soliton. Thus, we expect that the effect of
the latter manifests in the meson-baryon couplings via the single
quantity, $\rho^2$.

We can classify the cubic meson-nucleon-nucleon couplings from (\ref{5Deff})
into two classes. The first class consist of those whose leading large $N_c$
behaviors originate from the minimal coupling to ${\cal A}^{U(2)}$. This
includes dimensions-four couplings to vector mesons, and dimension-four
couplings to all iso-singlet mesons: vector, axial vector, or pseudo-scalar.
With the exception of those to iso-singlet vectors, this class of
couplings are subleading in $1/N_c$. Let us collectively denote them as
$g^{(I)}$.

The second class, $g^{(II)}$, have the leading $N_c$
contributions arising from the last term of (\ref{5Deff})
and thus proportional to $\rho^2$.
This class consists of coupling
to iso-triplet Goldstone boson (namely pions) \cite{Hong:2007kx},
minimal couplings to all iso-triplet axial vector mesons \cite{Hong:2007ay},
and dimension-five tensor couplings to all iso-triplet vector mesons
\cite{Kim:2009sr}. When we replace  $\rho^2_{\rm soliton}$ by $\rho^2$,
we therefore find
\begin{eqnarray}
g^{(I)}= g^{(I)}_{\rm soliton}\,,\qquad\qquad
g^{(II)}  = \sqrt{\frac54}\,g^{(II)}_{\rm soliton} \,.\quad
\end{eqnarray}
Numerically, the latter represents about 12\% increase for $g^{(II)}$'s.

Translating to quantities more directly related to data, we have
for example the charge form factor and the isospin-independent part
of nucleon-nucleon repulsive core unaffected, whereas the magnetic
form factor (thus the anomalous magnetic moments also) and the
isospin-dependent part of nucleon-nucleon potential would be increased
by factor of $5/4$ in the large $N_c$ limit.
See the next section for a direct matrix model computation of the
nucleon-nucleon potential at short distance, which confirms this
expectation.


Whether or not such a shift is  beneficial in
reproducing QCD with $N_c=3$ is unclear and needs to be studied more.
We would know the answer only after we have classified and computed
subleading corrections in this D4-D8 model. Regardless, we note that,
purely within the context of studying baryons in the D4-D8 model,
this represents a significant change to the leading $1/N_c$ computation
(and not a subleading correction) which could have not been obtained by any other method we know of.


\section{Two-body baryon interaction}
\label{sec4}

The baryon interaction at short distance can be obtained by classically
integrating out  $A_0$, as in the case of the single baryon. Now,
with two baryons, we have the matrices charged under the $U(2)$,
so, in particular, $A_0$ has four components,
\begin{eqnarray}
 A_0 = A_0^0 {\bf 1}_{2\times 2} +
A_0^1 \tau^1 + A_0^2 \tau^2 + A_0^3 \tau^3 \, .
\end{eqnarray}
Recall that, in the previous section,
solving for the equation of motion of
the overall $U(1)$ gauge field gave the
$1/\rho^2$ potential for individual baryon.
Now we are interested in integrating non-Abelian part of $A_0$ as well,
which should generate interaction energy for a pair of baryons.

Note that $A_0^a$ $(a=1,2,3)$ does not show up in our CS term in
(\ref{mm}), because the latter contains ${\rm tr} A_0$ only.
Terms including $A_0^a$ appear in the kinetic term of $X$ and that of $w$.

\subsection{Two-baryon configuration}

In order to evaluate the action,
first we fix the vacuum of the ADHM potential in (\ref{mm}).

\subsubsection{Single flavor}

The classical single baryon configuration
is specified by the vacuum configuration of $w$. In section
\ref{sec2.2}, we obtained the classical vacuum of the matrix
model as
\begin{eqnarray}
 w_{\dot{\alpha}}
= U\left(
\begin{array}{c}
 1\\0
\end{array}
\right)_{\!\!\!\dot\alpha}
\rho \,,
\end{eqnarray}
where $U$ is a $2\times 2$ unitary matrix. This representation is
equivalent to having $\rho_1$, $\rho_2$
in the notation of section \ref{sec2.2}.
This unitary matrix comes from the global symmetry (\ref{alphaglobal}).
Now we have two baryons, so the baryons are specified by
two unitary matrices,
\begin{eqnarray}
 w_{\dot{\alpha}}^{i=1}
= U^{(1)}\left(
\begin{array}{c}
 1\\0
\end{array}
\right)_{\!\!\!\dot\alpha}
\rho
\, ,
\quad
 w_{\dot{\alpha}}^{i=2}
= U^{(2)}\left(
\begin{array}{c}
 1\\0
\end{array}
\right)_{\!\!\!\dot\alpha}
\rho
\, .
\end{eqnarray}
One can work out the baryon interaction potential with this, but we will
perform the computation only in the 2-flavor case in the following,
since it is realistic.

\subsubsection{Two flavors}

For the two-flavor case,
the vacuum configuration of the potential is
given by the integration of the
$U(k)$-adjoint field $\vec{D}_{AB}$ in the matrix model (\ref{mm}),
which is the ADHM constraints,
\begin{eqnarray}
\vec{\tau}^{\;\dot{\alpha}}_{\;\;\;\dot{\beta}}
\left(
\bar{X}^{\dot{\beta}\alpha} X_{\alpha \dot{\alpha}}
+
\bar{w}^{\dot{\beta}}_{i} w_{\dot{\alpha}i} \right)_{BA}= 0\, .
\end{eqnarray}
Here we explicitly write the baryon index
$A,B = 1, ..., k$. In this section we treat
two instantons so $k = 2$.
The generic ADHM configuration satisfying this equation
is nothing but the ADHM data of two YM instantons.
It is given by
\begin{eqnarray}
&& X_M = \tau^3 \frac{r_M}{2} + \tau^1 Y_M \, ,
\label{configx}
\\
&& w^{A=1}_{\dot{\alpha}i}= U^{(A=1)}_{\dot\alpha i} \rho_1\, ,
\quad  w^{A=2}_{\dot{\alpha}i}= U^{(A=2)}_{\dot\alpha i} \rho_2\, ,
\label{configw}
\end{eqnarray}
where the locations of the two baryons are given by the diagonal entries
in $X_M$ so that $r_M$ is the inter-baryon distance, and
\begin{eqnarray}
&& Y_M \equiv -\frac{\rho_1\rho_2}{4(r_P)^2}{\rm tr}
\left[
\bar{\sigma}_M r_N \sigma_N
\left(
(U^{(1)})^\dagger U^{(2)}
-(U^{(2)})^\dagger U^{(1)}
\right)
\right] \, .
\label{defY}
\end{eqnarray}
Here $U^{(1)}$ and $U^{(2)}$ are  $SU(2)$ matrices which
denote the moduli parameters of each baryon, and
$\sigma_M\equiv (i\vec{\tau}, 1)$,
$\bar{\sigma}_M\equiv(-i\vec{\tau},1)$.

This ADHM data was
explicitly used in the soliton approach \cite{Hashimoto:2009ys}.
In terms of the YM
instanton, these degrees of freedom are gauge rotations of the ``flavor''
gauge group, and after the quantization, they become the spin and the
isospin of each baryon.
They can be written by
real unit vectors $a_M^{(1)}$ and $a_M^{(2)}$ as
\begin{eqnarray}
 U^{(1)} = i a^{(1)}_i \tau^i + a^{(1)}_4{\bf 1}_{2\times 2}\, , \quad
 U^{(2)} = i a^{(2)}_i \tau^i + a^{(2)}_4{\bf 1}_{2\times 2}\, , \quad
\end{eqnarray}
with $(a^{(1)}_4)^2 + (a^{(1)}_i)^2=1$,
$(a^{(2)}_4)^2 + (a^{(2)}_i)^2=1$. This is the correspondence to the
notation of \cite{Hata:2007mb}.
Using this, we obtain expressions which will be useful later,\footnote{
To obtain the first equality of (\ref{UUaa}), we have used the following
formula for unifying the double trace:
\begin{eqnarray}
\left({\rm tr}\left[\bar{\sigma}_M (b_0 + i b_i \tau^i)\right]\right) ^2
= 2 \; {\rm tr}\left[
\left(b^0 + i b^i \tau^i\right)
\left(b^0 - i b^j \tau^j\right)
\right].
\end{eqnarray}
Then, using $(a^{(1)}_4)^2 + (a^{(1)}_i)^2=1$ and
$(a^{(2)}_4)^2 + (a^{(2)}_i)^2=1$, the second equality of (\ref{UUaa})
follows.
}
\begin{eqnarray}
&& r_M Y_M = 0 \, ,
\label{ry}\\
&& Y_M Y_M = -\frac{\rho_1^2\rho_2^2}{8 (r_M)^2}
{\rm tr}
\left[
\left(U^{(1)\dagger} U^{(2)}-U^{(2)\dagger}U^{(1)}\right)^2
\right]
=
\frac{\rho_1^2\rho_2^2}{4 (r_M)^2}
\left(
1-(a_M^{(1)}a_M^{(2)})^2
\right)\, ,
\nonumber
\\
\label{UUaa}
\\
&& {\rm tr}\left[U^{(1)\dagger} U^{(2)}\right]
 =
{\rm tr}\left[U^{(2)\dagger} U^{(1)}\right]
=2 a_M^{(1)}a_M^{(2)}\, .
\end{eqnarray}

\subsection{Baryon interaction potential}

We shall integrate out the matrix $A_0$ and compute the two-baryon interaction
potential.
For integrating out the $SU(2)$ components of $A_0$,
we first write down
all the terms including those components in the matrix model action
(\ref{mm}). First,
\begin{eqnarray}
 {\rm tr}(D_0 X^M)^2
= 2\left(
(A_0^1)^2 r_M^2 + (A_0^2)^2 (r_M^2 + 4 Y_M^2)
+ 4 (A_0^3)^2 Y_M^2
\right) -8 A_0^1 A_0^3 r_M Y_M\, .
\label{dx2}
\end{eqnarray}
Here we used only (\ref{configx}). But if we further use the explicit
expression (\ref{defY}) and the vacuum for $w$ (\ref{configw}),
we obtain (\ref{ry}),
so the last term in (\ref{dx2}) vanishes.
Next, the $w$ kinetic term is
\begin{eqnarray}
 {\rm tr}D_0 \bar{w}^{\dot{\alpha}}_i D_0 w_{\dot{\alpha}i}
=
2(\rho_1^2+\rho_2^2)\left(
(A_0^0)^2 + (A_0^1)^2 + (A_0^2)^2+(A_0^3)^2
\right)
\nonumber \\
\vspace{20mm}
+ 4 \rho_1 \rho_2 A_0^0 A_0^1 \; {\rm tr}
\left[U^{(1)\dagger} U^{(2)}\right]
+ 4 (\rho_1^2-\rho_2^2)A_0^0 A_0^3\, .
\end{eqnarray}
In these kinetic terms,
the component $A_0^2$  appears only as a form $(A_0^2)^2$.
Thus we can minimize it independently with $A_0^2=0$, meaning that
we can just ignore the component $A_0^2$.
So, the total kinetic action plus the Chern-Simons term is
\begin{eqnarray}
&&
\frac{\lambda N_c M_{\rm KK}}{54 \pi}
\int\! dt \;
{\rm tr}\left[
(D_0 X^M)^2
+ D_0 \bar{w}^{\dot{\alpha}}_i D_0 w_{\dot{\alpha}i} \right]
+  N_c \int \! dt \; {\rm tr}A_0
\nonumber \\
&&
=
\frac{\lambda N_c M_{\rm KK}}{54 \pi}
\int\! dt \;
\biggm[
2(A_0^1)^2 r_M^2 + 8 (A_0^3)^2 Y_M^2
+
2(\rho_1^2+\rho_2^2)\left(
(A_0^0)^2 + (A_0^1)^2 +(A_0^3)^2
\right)
\nonumber
\\
&& \left.
\hspace{20mm}
+ 4 \rho_1 \rho_2 A_0^0 A_0^1 \; {\rm tr}
\left[U^{(1)\dagger} U^{(2)}\right]
+ 4 (\rho_1^2-\rho_2^2)A_0^0 A_0^3
+ \frac{108\pi}{\lambda M_{\rm KK}} A_0^0
\right]\, .
\end{eqnarray}
As the action is quadratic in the remaining components of $A_0$,
it is straightforward to integrate them out by diagonalizing the
interaction terms.
The resultant baryon interaction potential $V$ is determined from
$\int dt \; V= -S_{\rm on-shell}$ as
\begin{eqnarray}
 V =
\frac{27 \pi N_c}{\lambda M_{\rm KK}}\frac{1}{\rho_1^2 \rho_2^2}
\frac{((r_M)^2 + \rho_1^2 + \rho_2^2)
\left(4 (r_M)^2 (\rho_1^2+\rho_2^2)
- u \rho_1^2 \rho_2^2 \right)
}
{16((r_M)^2)^2 - 5 u(r_M)^2 (\rho_1^2\!+\!\rho_2^2)
- u
\left(\rho_1^4\! +\! \rho_2^4 - (u\!+\!2) \rho_1^2 \rho_2^2
\right)
}
\, ,
\end{eqnarray}
where
$u\equiv \left({\rm tr}\left[U^{(1)\dagger}U^{(2)}\right]\right)^2-4$.

In addition to this potential $V$,
we have another baryon interaction potential which comes from
the mass term of $X^4$ in the matrix model action (\ref{mm}). It is
easily evaluated as
\begin{eqnarray}
\frac{\lambda N_c M_{\rm KK}}{54 \pi}  \cdot \frac{2}{3} M_{\rm KK}^2
{\rm tr}(X^4)^2 =
\frac{\lambda N_c}{81 \pi} M_{\rm KK}^3
\left(
(r_4)^2/2 + 2(Y_4)^2
\right)\, .
\end{eqnarray}
Here $(r_4)^2$ term is the mass term in the single-baryon Hamiltonian,
so it does not contribute to the two-baryon interaction.
On the other hand, the off-diagonal element $Y^4$ is
intrinsically the interaction between the baryons. The definition
(\ref{defY}) is computed as
\begin{eqnarray}
 Y_4 = -\frac{\rho_1 \rho_2}{2 r_M^2}
r_i \; {\rm tr}
[i\tau^i (U^{(1)})^\dagger U^{(2)}]
\end{eqnarray}
where $i=1,2,3$, so
we can write the potential energy explicitly as
\begin{eqnarray}
 \frac{\lambda N_c M_{\rm KK}^3 }{162 \pi}
\left[(r_4)^2 +
\frac{\rho_1^2 \rho_2^2}{(r_M^2)^2} \left(
r_i \; {\rm tr}
\left[i\tau^i \left(U^{(1)}\right)^\dagger U^{(2)}\right]
\right)^2
\right]
\, .
\end{eqnarray}

Therefore, in total,
the two-baryon interaction Hamiltonian is given by
\begin{eqnarray}
V&=&
\frac{27 \pi N_c}{\lambda M_{\rm KK}}\frac{1}{\rho_1^2\rho_2^2}
\frac{((r_M)^2 + \rho_1^2 + \rho_2^2)
\left(4 (r_M)^2 (\rho_1^2+\rho_2^2)
- u \rho_1^2 \rho_2^2 \right)
}
{16((r_M)^2)^2 - 5 u(r_M)^2 (\rho_1^2+\rho_2^2)
- u
\left(\rho_1^4 + \rho_2^4 - (u+2) \rho_1^2 \rho_2^2
\right)
}
\nonumber \\
&&
+
 \frac{\lambda N_c M_{\rm KK}^3 }{162 \pi}
\frac{\rho_1^2 \rho_2^2}{(r_M^2)^2} \left(
r_j \; {\rm tr}
\left[i\tau^j \left(U^{(1)}\right)^\dagger U^{(2)}\right]
\right)^2
-
\frac{27\pi N_c}{4\lambda M_{\rm KK}}
\left(
\frac{1}{\rho_1^2} + \frac{1}{\rho_2^2}
\right)\, ,
\label{core}
\end{eqnarray}
where
$u\equiv \left({\rm tr}\left[U^{(1)\dagger}U^{(2)}\right]\right)^2-4$.
The last term is the subtraction of the single-baryon Hamiltonians
(the first term in (\ref{hamilours}), while
the second term (mass term) in (\ref{hamilours})
cancelled already).

Next, we shall evaluate this potential for a given quantum state of
the two baryons, and show that {\it
there is a universal repulsive core for any baryon
state.}

\subsection{Universal repulsive core}

First notice that in the classical limit, {\it i.e.}
$N_c \to \infty$, the expectation value of any function
$\langle f(\rho_1,\rho_2)\rangle$ with any given quantum state of the
baryons approaches
the classical value $f(\rho_1=\rho,\rho_2=\rho)$ where the classical
$\rho$ is given by (\ref{minrho}). The equivalence between $N_c
\to \infty$ limit and $\hbar \to 0$ limit is due to the fact that
our matrix action (\ref{mm}) has an overall factor $N_c$.
Since any deviation from the classical value is ${\cal O}(1/N_c^2)$,
while we are keeping only leading terms in the large $N_c$ expansion,
we in effect just need to put
$\rho_1 = \rho_2 (\equiv \rho)$ in our potential
(\ref{core}).\footnote{This argument is valid as long as the classical
value does not vanish and the wave function localizes for
$N_c\to \infty$,
which is in fact the present case for $\rho$ and $X^4$. As for the
spin/isospin encoded in $U^{(1)}$ and $U^{(2)}$ we cannot take the
classical value as the wave function is not localized in
the group space.}
We find
\begin{eqnarray}
V&=&
\frac{27 \pi N_c}{4\lambda M_{\rm KK}}
\frac{
\left({\rm tr}\left[U^{(1)\dagger}U^{(2)}\right]\right)^2
}{(r_M)^2 +2\rho^2 -\frac12
\left({\rm tr}\left[U^{(1)\dagger}U^{(2)}\right]\right)^2 \rho^2}
\nonumber
\\
& & \hspace{15mm}
+  \frac{\lambda N_c M_{\rm KK}^3 }{162 \pi}
\frac{\rho^4}{(r_M^2)^2} \left(
r_j \; {\rm tr}
\left[i\tau^j U^{(1)\dagger} U^{(2)}\right]
\right)^2\, .
\label{core3}
\end{eqnarray}
This is positive semi-definite.
In fact, we can show that our
potential (\ref{core3}) never vanish.
In order to see (\ref{core3}) never vanish, note that it could vanish if and only if
${\rm tr}\left[U^{(1)\dagger}U^{(2)}\right]=0$ and
${\rm tr}\bigl[i \tau^i U^{(1)\dagger} U^{(2)}\bigr]=0$
for any $i$.
However the latter condition implies
$U^{(1)\dagger} U^{(2)} \propto {\bf 1}_{2\times 2}$,
which contradicts with the former, therefore this is impossible.
In this way, we see that  (\ref{core3}) is {\it positive definite} and
as a result, there is {\it a universal repulsive potential (core) for
any choice of two-baryon quantum states}.

If we expand (\ref{core}) for $(r_M)^2 \gg \rho^2$,
we obtain a leading term
\begin{eqnarray}
V&=&
\frac{27\pi N_c}{64\lambda M_{\rm KK}} \frac{1}{(r_M)^2}
\left(
8\! +\! 6 \left({\rm tr}\left[U^{(1)\dagger}U^{(2)}\right]\right)^2
\!\!+
\left(\!-4\! +\!
5 \left({\rm tr}\left[U^{(1)\dagger}U^{(2)}\right]\right)^2
\right)
\!\left(\frac{\rho_2^2}{\rho_1^2}+ \frac{\rho_1^2}{\rho_2^2}\right)
\right)
\nonumber \\
&&
+
 \frac{\lambda N_c M_{\rm KK}^3 }{162 \pi}
\frac{\rho_1^2 \rho_2^2}{(r_M^2)^2} \left(
r_j \; {\rm tr}
\left[i\tau^j \left(U^{(1)}\right)^\dagger U^{(2)}\right]
\right)^2\, . \hspace{10mm}
\label{core2}
\end{eqnarray}
For the large $N_c$, again we can put $\rho_1=\rho_2=\rho$ and obtain
\begin{eqnarray}
V&=&
\frac{27 \pi N_c}{4\lambda M_{\rm KK}}
\left({\rm tr}\left[U^{(1)\dagger}U^{(2)}\right]\right)^2
\!\!
\frac{1}{(r_M)^2}
+
 \frac{\lambda N_c M_{\rm KK}^3 }{162 \pi}
\frac{\rho^4}{(r_M^2)^2} \left(
r_j \; {\rm tr}
\left[i\tau^j U^{(1)\dagger} U^{(2)}\right]
\right)^2\, .
\hspace{5mm}
\label{core4}
\end{eqnarray}
Again, we see the repulsive core.
The universal repulsive potential scales as $1/r^2$ at $r \gg \rho$
where $r$ is
the baryon separation.\footnote{Although the vector $r_M$ is in 4
spatial dimensions, once we take the VEV with the wave function of
$X^4$, it reduces to a 3-dimensional vector $r_i$ $(i=1,2,3)$.
This is again because
the leading term in the $1/N_c$ expansion is a classical value of $X^4$
which is zero.}

Let us consider the vacuum expectation value of the Hamiltonian
(\ref{core4}). The quantum state of the two baryons is specified by
$(\vec{I}_i, \vec{J}_i, n_\rho^{(1)}, n_Z^{(i)})$ with $i=1,2$ which
labels the two baryons. $\vec{I}$ ($\vec{J}$)
is the isospin (spin) of the baryon,
while $n_\rho$ and $n_Z$ are labels for excited baryon states
\cite{Hata:2007mb}. Explicit wave functions are given in
\cite{Hata:2007mb}. For nucleons $(|\vec{I}|=|\vec{J}|=1/2)$,
the spin/isospin wave functions are
\begin{eqnarray}
 \frac1{\pi} (\tau^2 U)_{IJ}
= \left(
\begin{array}{cc}
| p \uparrow \rangle & | p \downarrow \rangle
 \\
| n \uparrow \rangle & | n \downarrow \rangle
\end{array}
\right)_{IJ}
\end{eqnarray}
for each nucleon, so for our two-nucleon case the wave function is
\begin{eqnarray}
 \frac{1}{\pi^2}(\tau^2 U^{(1)})_{I_1J_1}
(\tau^2 U^{(2)})_{I_2J_2}
\end{eqnarray}
where $(I_1,J_1,I_2,J_2)$ are the third components of the
isospins and the spins for the nucleons, thus take values $\pm 1/2$.

Now we obtain an explicit repulsive core for nucleons.
At the leading order in $1/N_c$, we can simply take the classical
values for $\rho$ and $X^4$.
As for the spin/isospin part, we can use the formulas
given by (4.9), (4.11), and (4.18) of \cite{Hashimoto:2009ys},
\begin{eqnarray}
\left\langle
\left(
{\rm tr}\left[U^{(1)\dagger} U^{(2)}\right]
\right)^2
\right\rangle_{\!\!I_1,J_1,I_2,J_2} \hspace{-5mm}
&=& \; 1 + \frac{16}9 I_1^i I_2^i J_1^j J_2^j\, ,
\\
\left\langle
{\rm tr}\!\left[
i \tau^i U^{(1)\dagger} U^{(2)}
\right]
{\rm tr}\!\left[
i \tau^j U^{(1)\dagger} U^{(2)}
\right]
\right\rangle_{I_1,J_1,I_2,J_2} \hspace{-2mm}
&=& \delta^{ij} + \frac{16}{9}
I_1^k I_2^k
\left(
J_1^i J_2^j\! +\! J_2^j J_1^i\! -\! \delta^{ij} J_1^k J_2^k
\right) \, .
\nonumber
\end{eqnarray}
These formulas are for a given isospin $I_A^i$
and spin $J_A^i$ for the nucleon labeled as $A=1, 2$.
Finally, the vacuum expectation value of the potential (\ref{core4})
gives the central and the tensor forces,
\begin{eqnarray}
 \langle V\rangle_{I_1,J_1,I_2,J_2}
= V_{\rm C}(\vec{r}) + S_{12} V_{\rm T}(\vec{r})
\end{eqnarray}
with the standard definition
$S_{12} \equiv 12 J_1^i \hat{r}_i J_2^j \hat{r}_j - 4 J_1^i J_2^i$
(with $\hat{r}_i \equiv r_i/|r|$),
where
\begin{eqnarray}
 V_{\rm C}(\vec{r})
&=&
\pi \left(
\frac{3^3}{2} + 8 I_1^i I_2^i J_1^j J_2^j
\right)
\frac{N_c}{\lambda M_{\rm KK}}
\frac{1}{r^2} \, ,
\label{vc}
\\
 V_{\rm T}(\vec{r})
&=& 2\pi I_1^i I_2^i
\frac{N_c}{\lambda M_{\rm KK}}
\frac{1}{r^2} \, .
\label{vt}
\end{eqnarray}
This is the short-distance nuclear force obtained from our matrix model.
We find there is a repulsive core of nucleons.
The repulsive potential scales as
$1/r^2$ for the inter-nucleon distance $r$, which is a property peculiar
to the holographic model,
as noted in \cite{Hashimoto:2009ys}.\footnote{The nucleon-nucleon
potential
normally consists of two different regimes. One is the long distance
regime of order $1/M_{\rm KK}$ and beyond \cite{Kim:2009sr}, where the
four-dimensional exchanges of relatively
light mesons, including pions of course, dominate nucleon interaction and
induces attractive forces. The other is the repulsive core at short
distances. One might naively think that the latter is not too relevant
for formation of nuclei for example. However, nuclei are finely balanced
systems with very small binding energy, and at least numerically seem
sensitive to the details of the short distance repulsive core.}

Finally let us compare this result with the soliton approach
\cite{Hashimoto:2009ys} in which the two-baryon interaction potential
$V^{\rm soliton}(\vec{r})
= V_{\rm C}^{\rm soliton}(\vec{r}) + S_{12}
V_{\rm T}^{\rm soliton}(\vec{r})$
was computed as,
\begin{eqnarray}
V_{\rm C}^{\rm soliton}(\vec{r})
&=&
\pi \left(
\frac{3^3}{2} + \frac{32}{5}I_1^i I_2^i J_1^j J_2^j
\right)
\frac{N_c}{\lambda M_{\rm KK}}
\frac{1}{r^2} \, ,
\\
V_{\rm T}^{\rm soliton}(\vec{r})
&=& \frac{8\pi}{5}I_1^i I_2^i
\frac{N_c}{\lambda M_{\rm KK}}
\frac{1}{r^2} \, .
\end{eqnarray}
Compared to our (\ref{vc}) and (\ref{vt}), we can see that the structure of the
core is the same.\footnote{It is also
interesting to note that, at the stage of the Hamiltonian, the second term of (\ref{core})
completely coincides with what is called $H_1^{(SU(2))}$ in \cite{Hashimoto:2009ys} in its
structure.
This part basically gives the tensor force.
The difference in coefficients in the tensor forces
are due to the relation (\ref{classicalrhovalue}).
Nevertheless, since $H_1^{(SU(2))}$ came in \cite{Hashimoto:2009ys} from
a generalized
Osborn's formula, it is quite interesting that our matrix model can
reproduce quite easily the formula.} Furthermore, the numerical coefficients are only different
by factor 5/4 for the term proportional to $I_1^i I_2^i J_1^j J_2^j$ in $V_{\rm C}^{\rm soliton}(\vec{r})$
and for the term $V_{\rm T}^{\rm soliton}(\vec{r})$. One can see easily
that the difference in the last coefficient by factor 5/4 is due to
the relation (\ref{classicalrhovalue}). More generally, recall
that in the large $N_c$ one-boson exchange picture of nucleon-nucleon
potential \cite{Kim:2009sr}, the latter two isospin-dependent structures are
known to arise only from the second class of couplings, $g^{(II)}$, of
section \ref{mesonsection}. It is the coupling squared that enter the
potential, which explains the factor $5/4$ increase relative to the soliton model.
The isospin-independent central term arises, again in the large $N_c$,
from iso-singlet vector exchanges, couplings for which belong to the first
class, $g^{(I)}$, which explains the agreement.

It is encouraging that the soliton approach and our matrix approach
give qualitatively the same, and also quantitatively similar result
here as well. The large $N_c$ computation is
relatively stable in interpolating short distance and long distance
regime. At the same time, numerically the increase is hardly negligible
and may yet prove to be a big difference
in the end, when we have a good control over subleading $1/N_c$ and
$1/\lambda$ corrections in the model. Whether or not this will favor
the current D4-D8 model in simulating QCD is unclear for now, however,
since some of known subleading $1/N_c$ corrections appear at
comparable magnitude if $N_c=3$ is used and it is not known
how to catalog all such corrections.

Before closing, we would like to make comments on some of existing
computations of nucleon-nucleon potential. First of all, we again
emphasize that we only computed the repulsive core at very short
distance. As we mentioned above, this short-distance repulsive
core was computed by the solitonic methods in Ref.~\cite{Hashimoto:2009ys}
and also in Ref.~\cite{Kim:2008iy}, which differ from our result
only by a numerical factor. The fact that there is a universal $1/r^2$
behavior is a clear and unambiguous prediction of holography and
needs to be compared against other less speculative approaches,
possibly via lattice QCD simulations or by experiments.

Of more practical interest to most of nuclear community would be
the long-distance behavior, which should include an attractive
iso-singlet channel. From the holographic viewpoint of D4-D8 model,
this long-distance behavior was found in Ref.~\cite{Kim:2009sr} where
the authors computed one-boson exchange potential, with coupling constants
computed precisely using the solitonic picture of the baryon, and
indeed found the long-distance attractive potential with right
qualitative behaviors, such as allowing deuteron wavefunction
numerically \cite{ykim}. Our matrix model, in its current form, cannot
address this long distance behavior, unfortunately.

There exist extensive literatures on the matter of nucleon-nucleon
potential, which are too numerous and too diverse to list. See for
instance Ref.~\cite{Ericsson} for computations based on the chiral
perturbation theory. In the face of such extensive previous work, one
may ask why we should again try to compute nucleon-nucleon potential.
The first reason is that  at least the top-down models
such as ours have very few adjustable parameters and should be very
predictive and precise. This is not to say, of course, that the
latter would be more accurate. On the contrary, such top-down
holographic computations are less likely to be near the correct low
energy QCD physics.

What we aim to build is not the most phenomenologically
correct model, but rather those which can give us insights and results from top-down
approach and that cannot be given by any of conventional phenomenological methods.
The universal $1/r^2$ repulsive core above is one fine such example. The conventional
chiral perturbation, for instance, generates $1/r^3$ type repulsive
core from exchange rho mesons via tensor couplings as well as $1/r$
from omega meson exchange \cite{Ericsson}, whose precise behavior,
however, should be taken with a grain of salt at short distance below
the mass of the nucleon. In practice, one further introduces other
short-distance cut-offs to better simulate nature. Whether or not
the holographic $1/r^2$ repulsive core found by above holographic
computations represents a better solution to this short-distance
treatment of nuclear force and a more faithful image of real QCD
is a matter to be settled eventually by  lattice QCD simulations or by experiments.

\section{Conclusion}

We have derived a $U(k)$ matrix model
(\ref{mm}) which describes $k$-baryon systems at short
distance, by considering open string theory
on the wrapped baryon vertex D-branes embedded in the D4-D8
model of large $N_c$ holographic QCD. With this
matrix model, we computed the holographic size and the spectrum
of the baryon ($k=1$), and also short-distance
nuclear force ($k=2$). The latter  exhibits a repulsive core, which
has been quite important in nuclear and hadron physics for long years.

This model complements the Yang-Mills soliton picture of the holographic
baryon \cite{Hong:2007kx,Hata:2007mb,Hong:2007ay}, in that it is capable of
addressing short-distance behavior such as the crucial hadronic size estimate
of the baryon. Recall that the size found in these studies turned out to
be comparable to the (appropriately warped) string scale, which is too
small to be justified by the Yang-Mills method used there. Our new matrix
model is trustworthy well below the string scale, in the opposite end
of length scale, and represents an opportunity to check whether the
soliton picture gave us the correct estimates or not. As we summarize
shortly, the matrix model estimate gave us essentially the same
size up to a numerical factor of $(5/4)^{1/4}\sim 1$, showing that
stringy correction does very little to correct the soliton picture.
This is very fortunate for the underlying D4-D8 holographic QCD since, at least
for small number of baryons, the soliton picture has generated many
predictions that agreed with nuclear data \cite{Hong:2007ay,Hong:2007dq,Hashimoto:2008zw,Kim:2009sr}.
This is by far the best evidence we know of that justifies study of
baryons in holographic QCD, despite the latter's huge mass and
small size in the large $N_c$ and large $\lambda$ limit.

A distinguished character of our matrix model is its simplicity. In
particular, the form of the matrix model does not depend on the number
of the baryons described --- simply changing the rank of the matrix
allows one to treat larger number of baryons. By generalizing the
method we did in this paper, we can handle general $k$-body systems.
This is significant simplicity compared with the soliton picture, where
handling general $k$-solitons are quite painful due to the significant
increase of the number of
moduli parameters.

Therefore it is natural to apply our matrix model analysis to more
general $k$-body systems. In nuclear physics, the role of 3-body
forces is very important. For example, for few-body
nuclear bound states (light nuclei), the three-body forces are known to play a crucial
role. Three-body forces in the soliton picture in the D4-D8 system were
derived in \cite{Hashimoto:2009as}.
However, in that paper
the isospins and spins were treated classically due to the complications induced by
the increase of the number of moduli parameters, so the analysis was not
complete. Applying our matrix model, it is possible to treat generic
spins and isospins quantum-mechanically \cite{Hashimoto:2010ue}. It would be
interesting to see how the quantum spin and isospin effects
modify the previous soliton picture results \cite{Hashimoto:2009as}.

It would also be interesting to consider a large
$k$ limit, where the states are similar to extremely heavy nuclei
or a core of neutron stars. The reader may wonder the effect of
a back-reaction in that limit. Taking the large
$k$ limit certainly cause the back-reaction of the multiple D4'-branes in
D8-branes in D4-D8 system. However as we have seen, in our system, we
have no supersymmetric leftover, and as a result,
there is a repulsive forces between D4'-branes. So the D4'-branes cannot
approach closer than $\rho ={\cal O}(1/\sqrt{\lambda})
\sim l_s^{\rm eff}$,
which is too large compared with
$\rho \sim (l_s^{\rm eff})^2$ for Maldacena's decoupling limit \cite{Maldacena:1997re} in the
$l_s^{\rm eff} \to 0$
limit.\footnote{This argument could break down for finite
$l_s^{\rm eff}$ or in supersymmetric cases and we may consider a
holographic dual of the multi-baryon system in the $k \to \infty$
limit. It leads to ``holographic nuclei''
\cite{Hashimoto:2008jq,Hashimoto:2009pe}.}
Therefore the back-reaction is expected
not so strong enough to make a back-reacted gravitational throat.

Going back to the results for $k=1,2$ in this work, we note that
the results agree with the previous soliton approach qualitatively,
and in some case quantitatively also. The central isospin-independent
part of the repulsive core is one example. The holographic size squared
of the baryon also turned out to be larger only by a fraction
which, considering the vast separation of scales between the two
approaches by a factor of $\lambda^{1/2}$, seems pretty
innocuous. This tells us that physics of baryon in D4-D8 holographic QCD
is relatively stable against corrections. This stability of the
baryon physics against short distance corrections was previously
anticipated and argued for on the basis of an approximate supersymmetry
\cite{Yi:2009et}, and is borne out in our new matrix model approach.
Without such stability, study of baryons in holographic QCD would
have been very difficult and cumbersome.

On the other hand, the new results also show nontrivial change of
key observables, which cannot be ignored. From the viewpoint of low
energy effective theory in terms of mesons and baryons,
the change can be summarized in the large $N_c$ limit
as a universal and multiplicative
increase by $(5/4)^{1/2}$ of certain meson-nucleon-nucleon couplings.
These are couplings denoted as $g^{(II)}$ in section 3.5, and include
the leading axial couplings to pions, all isotriplet axial vector
couplings of the operator dimension four, and the dimension-five
derivative couplings to isotriplet vector mesons.  In particular,
the change came about not as subleading $1/N_c$ corrections, but
a change in coefficients of the leading terms.

The reader may wonder whether the matrix model can describe only
the short distance physics, such as the repulsive core of the nucleon.
As we argued in section 3.5 and reiterated above, the model is capable
of addressing the long-distance physics of soft-meson processes, via the
single quantity $\rho^2$, so in that indirect sense also tells us much
about baryons far away from one another. An entirely different question
is whether we can use the matrix model directly to compute interaction,
as we did in section 4, between well-separated baryons without
resorting to intermediate mesons. This is particularly important when
we wish to consider large $k$ physics where meson exchange viewpoint
would become quickly intractable.

The above matrix model does take into account  the dual warped geometry
of the pure QCD background, so it already knows about effect of glues
at energy scale $M_{\rm KK}$. Yet, how to emulate
the long-distance physics due to exchange of pions and other light
mesons directly by manipulating the current matrix model is a more
challenging problem. For example, in a similar problem of D0-brane interactions, it was
the one-loop effect in the open string side that emulated the long distance
graviton exchange between them. This agreement was explained by the
combination of supersymmetry and the worldsheet channel duality \cite{Banks:1996vh}.
In the present case, we must emulate an exchange of D8-D8 open strings
by a matrix model of D4'-D4' and D4'-D8 open strings. This appears
to be a qualitative different problem from the case of D0 interactions.
Absence of unbroken supersymmetry probably make matters a bit worse also.

Nevertheless, we believe that this can be achieved in the matrix model
with judicious insertions of effective operators and by considering
quantum effects. Once this is  done, on the other hand,
our matrix model may prove to be much more versatile than the
previous instanton soliton picture of baryons,
since it can be more easily generalized to many nucleon systems.
Nuclei with generic atomic number $k$ and perhaps other dense matter
systems would be more accessible that way.
Perhaps we can eventually construct a $U(k)$ matrix model, where
attractive long-distance forces and repulsive short-distance
forces are carefully balanced and produces,say, helium nucleus and carbon
nucleus.

In this context, we finally comment that we only made use of bosonic
fields in this work, even though there should be fermionic partners
in D4'-D8 matrix theory. For this work, we did not need loop computation,
which justifies this truncation. For a more complete matrix model
capable of directly dealing with baryons at large separation, we probably
need to consider fermionic fields as well, which is beyond the scope of
this paper.


Eventually we hope to be able to compute other interesting quantities in
our matrix model, such as currents associated with chiral symmetry,
energy-momentum tensor and equations of states from that. In particular,
equations of states with large baryon number $k$ are important to understand the
dynamics of astrophysical compact starts, such as neutron stars.

\acknowledgments
We would like to thank Yukawa Institute in Kyoto university and the
organizers of the workshop on Branes, Strings and Black Holes, where
this project started. K.H.~is grateful to D.~Tong for a discussion,
and would like to thank DAMTP in university of
Cambridge, National Taiwan Normal
university, IPMU in university of
Tokyo and CERN for hospitality. N.I.~thanks RIKEN for hospitality.
P.Y. was supported in part by the National Research Foundation of Korea (NRF)
funded by the Ministry of Education, Science and Technology via the Center for
Quantum Spacetime (grant number 2005-0049409),
Basic Science Research Program (grant number 2010-0013526),
and Basic Research Promotion Fund (grant number KRF-2007-314-C00052) .


\end{document}